\documentclass[twocolumn,pre,floats,superscriptaddress]{revtex4}

\usepackage{graphicx}
\usepackage{amssymb}
\usepackage{amsmath,bm}
\usepackage{psfrag}
\usepackage{epsfig}
\usepackage{float}
\usepackage{color}
\pdfoutput=1

\begin{document}
\title{Model for computing kinetics of the graphene edge epitaxial growth on copper}

\author{Mikhail Khenner}
\affiliation{Department of Mathematics, Western Kentucky University, Bowling Green, KY 42101}
\affiliation{Applied Physics Institute, Western Kentucky University, Bowling Green, KY 42101}

\begin{abstract}

A basic kinetic model that incorporates a coupled dynamics of the carbon atoms and dimers on a copper surface is used to compute growth of a single-layer 
graphene island. The speed of the island's edge advancement on Cu[111] and Cu[100] surfaces is computed as a function of the growth temperature and pressure.
Spatially resolved concentration profiles of the atoms and dimers are determined, and the contributions provided by these species to the growth speed are discussed.
Island growth in the conditions of a thermal cycling is studied.

\end{abstract}

\date{\today}
\maketitle

\begin{center}
\small{{\it Journal information: Physical Review E {\bf 93}, 062806 (2016); DOI: 10.1103/PhysRevE.93.062806}}
\end{center}


\section{Introduction}

Epitaxial growth of a high quality, large area single- and multi-layer graphene sheets on a transition-metal substrates is presently in the focus of the research efforts worldwide, as has been discussed in several review articles 
\cite{Mattevi,Reina,Tetlow}. 
Strategies were developed to grow the graphene sheets with the area of up to 1 cm$^2$ using the chemical vapor deposition (CVD) of a hydrocarbons (such as methane, CH$_4$) on copper, an abundant and inexpensive substrate \cite{Li,Yan,Wang}. Alongside the experimental efforts, modeling of the CVD graphene growth on Cu was also attempted. These studies can be divided into three groups: \textit{ab initio} and Kinetic Monte Carlo (KMC) methods \cite{Wu,Wu1,Gaillard}, rate equations \cite{Wu,ZV}, and phase-field methods \cite{LS}.
By assuming the anisotropic diffusion on Cu of the carbon atoms and their anisotropic attachment
to the islands, the authors of the latter reference succeeded in computing the growth of the multi-lobe graphene islands on the substrates of different crystallographic orientations. However, 
with the focus of the study on the morphologies, the kinetic maps of the growth rate dependence on the controllable process parameters, such as the pressure and temperature, were not computed.

In this paper we describe a simpler, one-dimensional PDE model, whose purpose is to compute the growth rate of a single graphene island as a function of three control parameters: the crystallographic orientation of the substrate, 
the growth temperature,
and the pressure of a gas of the carbon atoms that impinge on the substrate. In the manner of Ref. \cite{LS},  we abstract from the details of the dissociation of a hydrocarbons, assuming that it results
in the  carbon gas from which the carbon atoms are adsorbed on Cu surface. However, we recognize, as is pointed out in the \textit{ab initio} studies, that besides the carbon atoms there is other
diffusing species that may contribute to the island growth \cite{Tetlow} - of which the carbon dimers are thought to be the most important \cite{Wu,Celebi,Wu1}. 
Our hybrid model can be seen as an extension, directly informed by the activation energies from the \textit{ab initio} calculations \cite{Wu}, of the classical BCF-type modeling \cite{BCF} to two 
interacting and diffusing species that feed growth of the graphene island edge. This results in a coupled PDE problem for the concentration fields on the substrate. Both PDEs are also coupled, through the boundary conditions,
to an ODE for the position of the island edge.

\section{Model Formulation}
\label{ModForm}

%

\begin{figure}[H]
\centering
\includegraphics[width=3.5in]{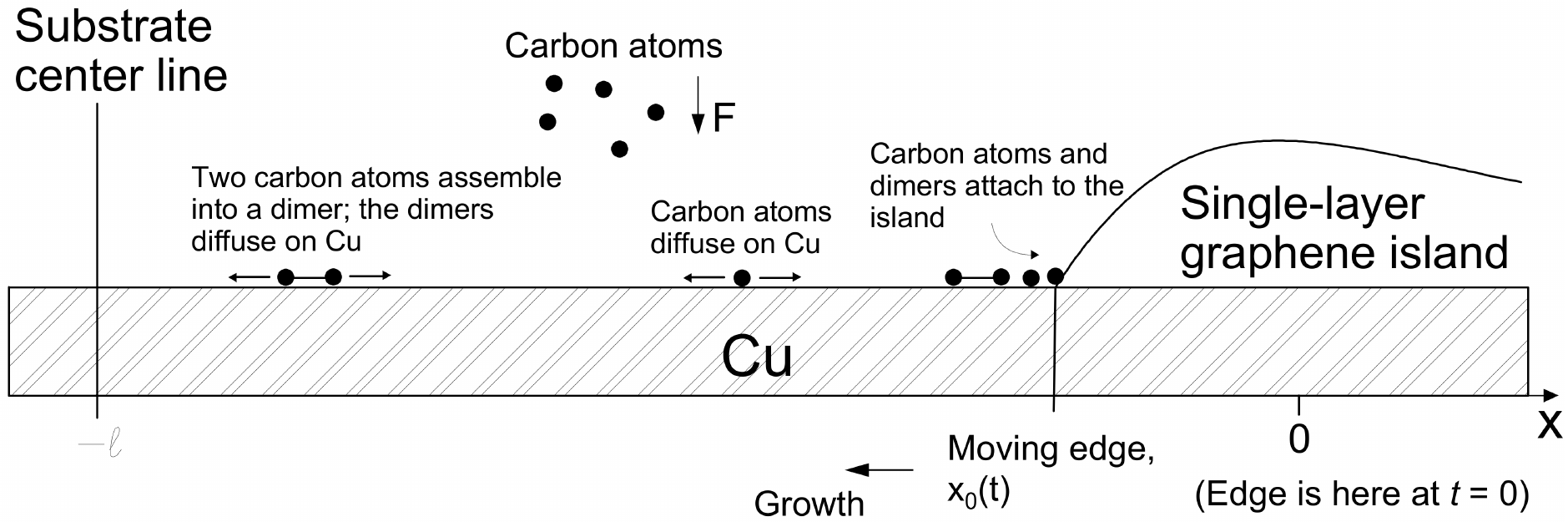}
\caption{Sketch of the graphene island growing on Cu, with the atomic events shown.
}
\label{Geom}
\end{figure}

As we pointed out in the Introduction, the model is aimed at computing the velocity, $\dot x_0(t)$, of a growing edge of a single-layer graphene island.
Here $x_0(t)$ is the position of the edge, see Fig. \ref{Geom}.

Since the edge grows predominantly by attachment of the carbon atoms and dimers \cite{Wu,Celebi,Wu1}, let
$C(x,t)$ and $C_2(x,t)$ be the concentrations of the carbon atoms adsorbed on Cu and the carbon dimers, respectively.
The latter result from the assembly of two previously adsorbed carbon atoms.

The model is comprised of the following PDEs and boundary conditions \cite{RefToZV}.

\begin{itemize}
\item Evolution equation for the concentration $C$ on the section of the copper substrate that is not yet covered by the growing graphene island:
\begin{equation}
\frac{\partial C}{\partial t} = D_c \frac{\partial^2 C}{\partial x^2} - \chi_c C^2 + F, \; -\ell \le x \le x_0(t).
\label{C_eq}
\end{equation}
Here $D_c$ is the carbon atoms diffusivity, $F$ the adsorption flux, 
and the sink term $-\chi_c C^2$ describes the loss of the carbon atoms due to their assembly into the dimers; the kinetics of this loss is 
reciprocal in $t$, e.g.
$C\sim 1/\chi_c t$, as follows from the ODE $dC/dt =  - \chi_c C^2$. We found that it is not necessary to 
include atoms desorption in Eq. (\ref{C_eq}), particularly since the desorption rate has not been published and because even without the desorption
the concentration is quite small (Fig. \ref{Concentrations}). Eq. (\ref{C_eq}) is a well-posed nonlinear PDE with a unique solution for all $t>0$ \cite{L,V}.

The boundary conditions  for $C$ are:
\begin{eqnarray}
x&=& - \ell:\quad \frac{\partial C}{\partial x} = 0, \label{bc1} \\
x&=&x_0(t): \quad - D_c \frac{\partial C}{\partial x} = K_c \left(C - C_{eq}\right). \label{bc2}
\end{eqnarray}
The first boundary condition states that far from the graphene island (at the center of the substrate) the carbon concentration profile is symmetric. 
The second one states that at the growing edge the flux of the carbon atoms
is proportional to the difference between the concentration there and the equilibrium
concentration \cite{Uwaha}; the proportionality parameter $K_c$ is the kinetic coefficient, which gives a measure of the ease with
which the carbon atoms can attach to the edge.

\item Evolution equation for the concentration $C_2$, also on the section of the substrate not yet covered by the growing graphene island:
\begin{eqnarray}
\frac{\partial C_2}{\partial t} & = & D_{c2} \frac{\partial^2 C_2}{\partial x^2} - \kappa_{c2} C_2 + \chi_c C^2, \label{C_2_eq} \\
& & -\ell \le x \le x_0(t). \nonumber
\end{eqnarray}
Here $D_{c2}$ is the diffusivity of the carbon dimers, $\kappa_{c2}$ their desorption rate,  
and the source term $\chi_c C^2$ is due to assembly of the carbon atoms into dimers. Notice that through this term the linear Eq. (\ref{C_2_eq}) is one-way coupled to Eq. (\ref{C_eq}).

The boundary conditions for $C_2$ mirror those for $C$:
\begin{eqnarray}
x&=& - \ell:\quad \frac{\partial C_2}{\partial x} = 0, \label{bc3} \\
x&=&x_0(t): \quad -D_{c2} \frac{\partial C_2}{\partial x} = K_{c2} \left(C_2 - C_{eq2}\right). \label{bc4}
\end{eqnarray}
Here $K_{c2}$ is the attachment coefficient of the dimers and $C_{eq2}$ the equilibrium concentration for the growth by the dimers attachment.

\item Equation of the edge growth:
\begin{eqnarray}
\dot x_0(t) & = & -\Omega K_c \left[C(x_0(t),t) - C_{eq}\right] - \label{x0_eq} \\
& & 2\Omega K_{c2}\left[C_2(x_0(t),t) - C_{eq2}\right],\; x_0(0)=0. \nonumber
\end{eqnarray}
This equation states that the edge velocity is the sum of the contributions resulting from the attachment of the atoms and dimers, where each contribution is proportional to 
the deviation at the edge of the corresponding concentration from its equilibrium value \cite{Uwaha}. 
$\Omega= \pi a^2$ is the atomic area, where $a=7\times 10^{-9}$ cm is the radius of the carbon atom.

\end{itemize}

\vspace{-0.5cm}
Equations (\ref{C_eq})-(\ref{x0_eq}) are made dimensionless by choosing $\ell$, $\ell^2/D_c$ and $1/\Omega$ as the length, time, and concentration scale, respectively.
Keeping the same notations for the dimensionless variables, the dimensionless system reads:
\begin{eqnarray}
\frac{\partial C}{\partial t} &=& \frac{\partial^2 C}{\partial x^2} - \alpha C^2 + \beta,  \label{C_eq_a}\\
\frac{\partial C_2}{\partial t} &=& D \frac{\partial^2 C_2}{\partial x^2} - \delta C_2 + \alpha C^2,  \label{C_2_eq_a}\\
\dot x_0(t) &=&  -R_c \left[C(x_0(t),t) - \Omega C_{eq}\right] - \label{x0_eq_a}\\ 
& & 2 R_{c2} D\left[C_2(x_0(t),t) - \Omega C_{eq2}\right], \; x_0(0)=0. \nonumber \\
x&=& -1:\quad \frac{\partial C}{\partial x} = 0,\; \frac{\partial C_2}{\partial x} = 0; \label{bc1a} \\
x&=&x_0(t): \quad \frac{\partial C}{\partial x} = R_c \left(\Omega C_{eq} - C\right), \label{bc2a} \\ 
& & \hspace{1.4cm} \frac{\partial C_2}{\partial x} =  R_{c2} \left(\Omega C_{eq2} - C_2\right). \nonumber
\end{eqnarray}
Here the eight parameters are: $\alpha = \chi_c \ell^2/D_c$ (the assembly rate of the atoms into the dimers), $\beta = F \Omega \ell^2/D_c$ (the adsorption flux of the atoms), 
$\delta = \kappa_{c2} \ell^2/D_c$ (the desorption rate of the dimers), 
$D = D_{c2}/D_c$ (the ratio of the diffusivities), $R_c= K_c \ell/D_c$ (the attachment rate of the atoms), $R_{c2}= K_{c2} \ell/D_{c2}$ (the attachment rate of the dimers), $\Omega C_{eq}$, and $\Omega C_{eq2}$
(the dimensionless equilibrium concentrations).

The initial condition for $C$ is taken in the form of a smoothed step function with a narrow transition,
in the middle of the interval, from a smaller positive value at $x=-\ell$ to a larger value at 
$x=x_0(0)=0$. Zero initial condition for $C_2$ is taken, e.g. at $t=0$ there is no dimers on the substrate.

Apart from the multiple parameters, the system (\ref{C_eq_a})-(\ref{bc2a}) looks deceptively simple. However, this is the moving-boundary problem, since the position $x_0(t)$ of the graphene edge is \textit{a priori}
unknown and must be found alongside with the concentrations. Due to a moving edge, any change in the concentrations gradients near the edge affects the edge growth speed, and the change in speed
in turn affects the concentrations near the edge and beyond. After focusing on the physical parameters in the next section, in Section \ref{NumMeth} the procedure for the numerical solution of this system of equations is described.

\section{Physical parameters}
\label{PhysPara}

All physical parameters are taken in the Arrhenius form, with the most recent and complete, 
to our knowledge, values of the activation energies \cite{Wu} (see Table I).  The pre-exponential factors
are taken proportional to $k_B T/h$, where $h$ is Planck's constant \cite{ZV}.
\begin{eqnarray}
D_c &=& \frac{k_B T a^2}{h} e^{-E_{Dc}/k_B T},\; D_{c2} = \frac{k_B T a^2}{h} e^{-E_{Dc2}/k_B T}, \nonumber \\ 
\chi_c & =& \frac{k_B T}{h} e^{-E_{\chi}/k_B T},\; F =\frac{P_0}{\sqrt{2 \pi m k_B T}}e^{-E_{ad}/k_B T}, \nonumber \\ 
\kappa_{c2} & =& \frac{k_B T}{h} e^{-E_{\kappa}/k_B T}, \nonumber \\
K_c &=& \frac{k_B T a}{h} e^{-E_{Kc}/k_B T},\ K_{c2} = \frac{k_B T a}{h} e^{-E_{Kc2}/k_B T}, \nonumber \\
C_{eq} &=& \Omega^{-1} e^{-E_{Ceq}/k_B T},\; C_{eq2} = \Omega^{-1} e^{-E_{Ceq2}/k_B T}.
\label{params}
\end{eqnarray}
\begin{center}
    \begin{tabular}{| c | c | c | c | c | c | c | c| c | c |}
    \hline
    \small{Cu surface} &  \small{$E_{Dc}$} & \small{$E_{Dc2}$} & \small{$E_{\chi}$} & \small{$E_{\kappa}$} & \small{$E_{Kc}$} & \small{$E_{Kc2}$} & \small{$E_{ad}$} & 
\small{$E_{Ceq}$} & \small{$E_{Ceq2}$}
\\ \hline
    \small{[111]} & \small{0.5} & \small{0.49} & \small{0.9} & \small{1.7} & \small{0.71} & \small{0.74} &  \small{0.1} & \small{0.87} & \small{0.87}\\ \hline
    \small{[100]} & \small{1.11} & \small{0.62} & \small{0.59} & \small{1.7} & \small{1.42} & \small{1.07} &  \small{0.1} &  \small{0.87} & \small{0.87}\\
    \hline
    \end{tabular}\vspace{0.5cm}\\
Table 1. Activation energies (in eV).
\end{center}

Values for $E_{\kappa},\; E_{ad},\; E_{Ceq}$ and $E_{Ceq2}$ were not published for graphene growth on copper. Thus in the Table 1 we adopt the generic values for $E_{ad},\; E_{Ceq}$ and $E_{Ceq2}$ 
\cite{Hong,Schulze},
and for $E_{\kappa}$ we adopt a value that partially curtails the otherwise unlimited growth of the dimers concentration 
(caused by the perpetual assembly of the carbon atoms into dimers), allowing the computation to proceed until
the edge grows over the entire available substrate. This value is large, thus the desorption flux is small.

Carbon gas pressure $P_0$ is varied in the range 100$-$600 mTorr, $m=2\times 10^{-23}$ g
is the molecular weight of carbon, the temperature $T$ is in the interval 873$-$1273 K, and the half-width
of the substrate $\ell = 1$ mm.

\section{Solution methods}
\label{NumMeth}

Since solving PDEs on a time-dependent domains is difficult, we first map the interval $-1\le x\le x_0(t)$ onto a fixed interval $-1\le \xi\le 0$ for the new space variable $\xi$,
using the transformation 
\begin{eqnarray}
\xi & = & \frac{x-x_0(t)}{1+x_0(t)},\; C(x,t) = U(\xi(x,t),t), \label{transform1} \\ 
& & C_2(x,t) = V(\xi(x,t),t), \nonumber
\end{eqnarray}
where $U$ and $V$ are the concentrations of the atoms and dimers on the fixed interval. 
Then the system (\ref{C_eq_a})-(\ref{bc2a}) takes the form:
\begin{eqnarray}
\frac{\partial U}{\partial t} &=& \left(\frac{1}{1+x_0(t)}\right)^2\frac{\partial^2 U}{\partial \xi^2} + \label{C_eq_b}\\ 
& & \dot x_0(t)\left(\frac{1+\xi}{1+x_0(t)}\right)\frac{\partial U}{\partial \xi} - \alpha U^2 + \beta, \nonumber \\
\frac{\partial V}{\partial t} &=& D \left(\frac{1}{1+x_0(t)}\right)^2\frac{\partial^2 V}{\partial \xi^2} + \label{C_2_eq_b}\\ 
& & \dot x_0(t)\left(\frac{1+\xi}{1+x_0(t)}\right)\frac{\partial V}{\partial \xi} - \delta V + \alpha U^2, \nonumber \\
\dot x_0(t) &=&  -R_c \left[U(0,t) - \Omega C_{eq}\right] - \label{x0_eq_b}\\ 
& & 2 R_{c2} D\left[V(0,t) - \Omega C_{eq2}\right],\quad x_0(0)=0. \nonumber \\
\xi&=& -1:\quad \frac{\partial U}{\partial \xi} = 0,\; \frac{\partial V}{\partial \xi} = 0; \label{bc1b} \\
\xi&=&0: \quad \frac{\partial U}{\partial \xi} = \left(1+x_0(t)\right)R_c \left(\Omega C_{eq} - U \right),\label{bc2b} \\ 
& & \hspace{0.9cm} \frac{\partial V}{\partial \xi} = \left(1+x_0(t)\right)R_{c2} \left(\Omega C_{eq2} - V \right). \nonumber
\end{eqnarray}
\begin{figure}[H]
\centering
\includegraphics[width=2.0in]{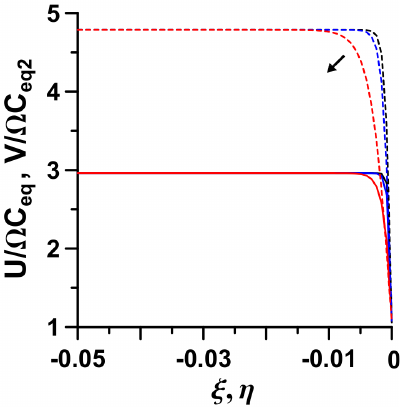}
\caption{(Color online). Example dimensionless concentrations of the atoms (solid lines) and dimers (dashed lines), shown in units of the respective dimensionless equilibrium concentration, 
vs. the transformed dimensionless coordinate along the Cu[111] substrate (see Sec. \ref{NumMeth}). $\xi=0$ (or $\eta=0$) corresponds to the growing edge. Time increases in the direction shown by the arrow (from the black to the red). Notice that the concentrations are not fixed at $\xi=0$ (by the boundary conditions (\ref{bc2b})). Also notice that the dimers concentration is larger than the one of the atoms, which
corroborates the findings in the \textit{ab initio} computations \cite{Wu}.
}
\label{Concentrations}
\end{figure}

In the transformed system the edge of the graphene island is at $\xi=0$ at all times. However, the true edge position $x_0(t)$ is found from Eq. (\ref{x0_eq_b}) and therefore the kinetics of growth is preserved.  
The system (\ref{C_eq_b})-(\ref{bc2b}) is the initial-boundary value problem for two one-way coupled PDEs (with the variable coefficients), which are also coupled to the first-order ODE
for $x_0(t)$. For the solution of this systems we adopted the classical Method of Lines (MOL), which converts the PDEs into ODEs by discretizing the space variable using finite 
differences. However, with the realistic physical parameters 
from Sec. \ref{PhysPara}  the computed concentration profiles
feature a steep boundary layer at the growing edge (see Fig. \ref{Concentrations}, also Figures \ref{Concentrations_T=1273KtoT=973KtoT=1273K_[111]}(a,b)). For the prediction of the edge growth rate, it is crucial to 
resolve these layers with a high accuracy. We found that this is achieved by a method that we describe
below.

First, the space variable $\xi$ is transformed as \cite{SpotzCarey}:
\begin{equation}
\xi(\eta) = \eta +\frac{\gamma}{\pi}\sin{\pi \eta},\; -1\le \eta\le 0,\; |\gamma| < 1.
\label{xi_eta}
\end{equation}
Notice that this map is invertible when the absolute value of the parameter $\gamma$ is less than one, 
also $\eta(0)=0,\; \eta(-1)=-1$. The purpose of the 
transformation is to map the would-be non-uniform computational grid on $-1\le \xi\le 0$ (where at $\gamma < 0$ 
the grid points are clustered
near $\xi=0$) onto a uniform grid on $-1\le \eta\le 0$. 
In all computations we used $\gamma = -0.95$.

Next, the final transformed system is discretized in $\eta$ using the second-order finite differences with the fixed grid spacings $h$ and $h/2$, and two ODE systems resulting from such discretizations are solved 
independently and in parallel using the same initial condition. Richardson interpolation is performed after a
fixed number of time steps. This way a spatially fourth-order accurate solution is obtained
on a coarse grid. This solution is then interpolated onto a fine grid before the next step is taken. The temporal accuracy
is achieved automatically by an ODE solver. The grid refinement study was performed, 
which indicated that using $h=0.0008$ results in the needed overall computational accuracy for all parameters
values of interest. 

\section{Results}

\begin{figure}[H]
\centering
\includegraphics[width=3.25in]{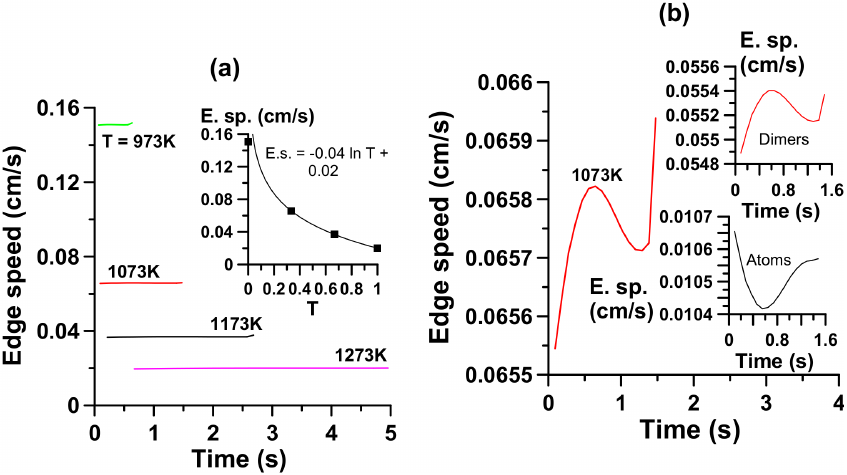}
\caption{(Color online.) Cu[111] surface. (a) Edge speed, $|x_0(t)|$, vs. the time at $P_0=500$ mTorr and various temperatures. The last point on each curve corresponds to the time 
$t_{final}$ at which the edge reaches the midpoint
of the substrate, $x=-\ell=-0.1$ cm. Inset shows the mean edge speed vs. the temperature (that was first mapped onto the unit interval), where the mean is calculated over the time
interval from zero to $t_{final}$ for each curve; the line is the logarithmic fit to the data shown by squares. (b) Magnification of the $T=1073$ K curve from the panel (a). 
Insets show separately the components of this speed due to the attachment to the edge of the atoms and dimers.
}
\label{Speed_vs_t_and_Speed_vs_T}
\end{figure}

\begin{figure}[H]
\centering
\includegraphics[width=2.0in]{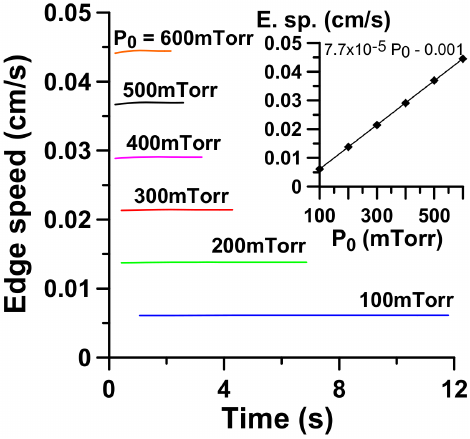}
\caption{(Color online). Cu[111] surface. Edge speed vs. the time at $T=1173$ K and various pressures. Inset: the mean speed vs. the pressure (squares) and the linear fit, also
at $T=1173$ K.
}
\label{Speed_vs_P0}
\end{figure}

\begin{figure}[H]
\centering
\includegraphics[width=3.25in]{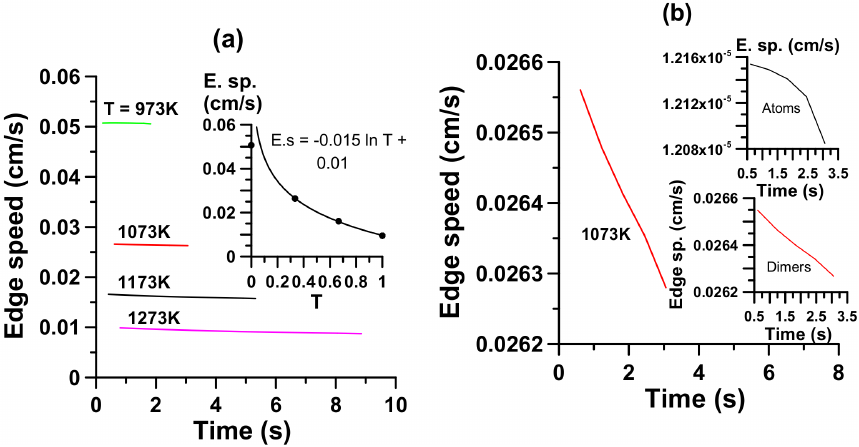}
\caption{(Color online.) Same as Fig. \ref{Speed_vs_t_and_Speed_vs_T}, but for Cu[100] surface.
}
\label{Speed_vs_t_and_Speed_vs_T_[100]}
\end{figure}

\begin{figure}[H]
\centering
\includegraphics[width=2.0in]{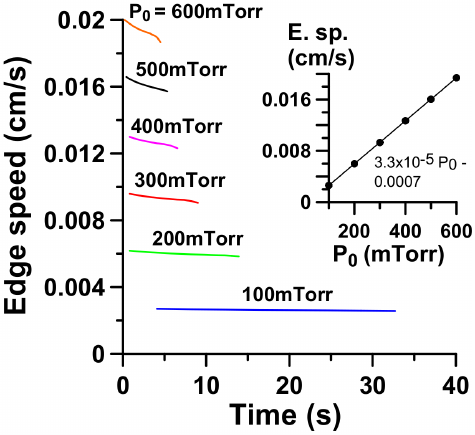}
\caption{(Color online). Same as Fig. \ref{Speed_vs_P0}, but for Cu[100] surface. 
}
\label{Speed_vs_P0_[100]}
\end{figure}

We begin this section with the comparisons of Figures \ref{Speed_vs_t_and_Speed_vs_T} and \ref{Speed_vs_P0}, computed for graphene growth on Cu[111] surface, with the corresponding Figures
\ref{Speed_vs_t_and_Speed_vs_T_[100]} and \ref{Speed_vs_P0_[100]} for the growth on Cu[100] surface.

In Figures \ref{Speed_vs_t_and_Speed_vs_T}(a) and \ref{Speed_vs_t_and_Speed_vs_T_[100]}(a) it can be seen that 
the growth slows down as the temperature increases, which perhaps explains better graphene quality and larger islands at higher growth temperatures \cite{Celebi,Li,Yan,Mattevi,Wang}.
From the insets to these Figures, it appears that the slow-down is logarithmic. This is a new and important model prediction, as the quantitative experimental results on the growth speed scaling
with the temperature have not been published.
At each temperature the speed is nearly a constant value for the entire duration of the simulation (changing less than 1\%, see Figures \ref{Speed_vs_t_and_Speed_vs_T}(b) and \ref{Speed_vs_t_and_Speed_vs_T_[100]}(b)). Also we noticed that the speed is smaller on Cu[100] and it slowly and monotonically decreases with time on this surface, while on 
Cu[111] surface the curve is S-shaped; the latter dynamics is somewhat similar to the one shown in Fig. 2 of Ref. \cite{Li}.
Attachment of the dimers provides the major contribution to the growth speed (see the insets). In the case of growth on Cu[111], the contribution from the dimers exceeds by a factor of five 
the one from the atoms; on Cu[100], the atoms provide a negligible contribution.
This supports the recent conclusions in the \textit{ab initio} \cite{Wu,Wu1} and experimental papers \cite{Celebi,Li,Yan} that the graphene edge grows primarily by the dimers attachment. 

Figures \ref{Speed_vs_P0} and \ref{Speed_vs_P0_[100]} show the dependencies of the edge speed on the time and pressure at a fixed temperature. The speed increases linearly with $P_0$.
This is another key model prediction that remains to be supported by the experiment; the quantitative experimental data were not published.
We remark here that the computed growth speeds shown in Figures \ref{Speed_vs_t_and_Speed_vs_T} - \ref{Speed_vs_P0_[100]} exceed by a few orders of magnitude the 
speeds that are reported in the experimental papers.
Values from the experiments 
seem to be of the order $10^{-6} - 10^{-5}$ cm$/$s for the temperature range 
that we use in the computations. We conjecture that the discrepancies are primarily due to the larger $P_0$ values used in our computations than the carbon partial pressures in the experiments, as well as because the adopted $E_{ad}$ value is approximate.
Since the pressures of the hydrocarbons or the evaporated carbon are not consistently reported in the experimental literature, 
we took for $P_0$ the set of ``growth pressure" (or ``chamber pressure") values from Ref. \cite{Mattevi}.  From the inset of Fig. \ref{Speed_vs_P0} 
one can see that the speed of the order $10^{-5}$ cm$/$s 
would result when the fit is extrapolated to $P_0\sim 13$ mTorr (for Cu[111] surface, see Fig. \ref{Speed_vs_P0_[100]}, this value is 21 mTorr).
These extrapolated values are order-of-magnitude consistent with those reported in Refs. \cite{Celebi,Li} in conjunction with the growth rates of the orders that we stated above.

\begin{figure}[H]
\vspace{-0.3cm}
\centering
\includegraphics[width=1.75in]{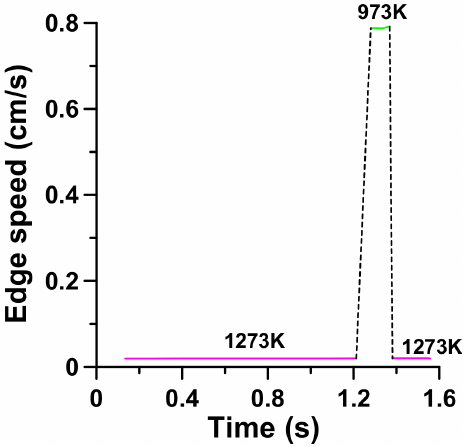}
\vspace{-0.1cm}
\caption{(Color online). Cu[111] surface, $P_0=500$ mTorr. The temperature is quenched from 1273 K to 973 K and back. 
}
\label{Speed_vs_t_T=1273KtoT=973KtoT=1273K_[111]}
\end{figure}

\begin{figure}[H]
\vspace{-0.7cm}
\centering
\includegraphics[width=3.25in]{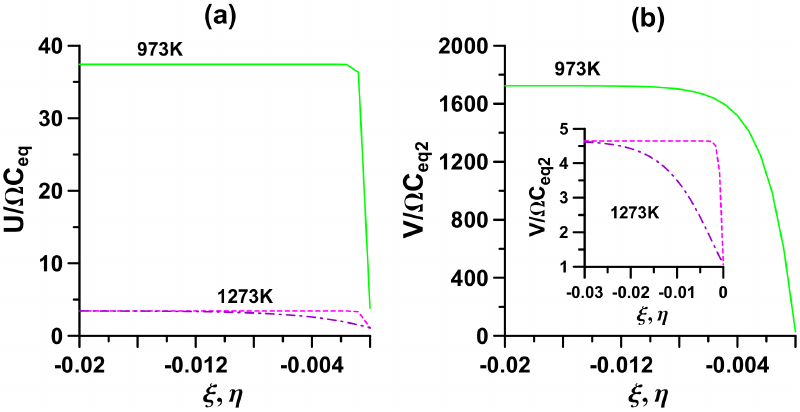} 
\caption{(Color online.) Cu[111] surface, $P_0=500$ mTorr. (a) Atoms concentrations at $T=1273$ K and 973 K before the cool-down (dashed magenta line), after the cool-down 
(solid green line) and after the warm-up (dash-dotted purple line).  (b) Dimers concentrations; the lines coloring is the same as in (a).
}
\label{Concentrations_T=1273KtoT=973KtoT=1273K_[111]}
\end{figure}

It is common in the experiments to employ thermal cycles during growth or sharply decrease the temperature at the very end of the growth phase.  
This typically results in better quality of the graphene layer, also its area is enlarged \cite{Yan,Gao,Koos}. Why this happens is not well understood \cite{Kim}.  
Our model is well-suited for giving some insights into this situation. We started the computation using the parameters at 1273 K and computed for some time, then 
instantaneously switched to the parameters at 973 K and computed more, and finally switched back to the parameters at 1273 K and computed until the substrate overgrowth
by a graphene sheet was completed. 
In Fig. \ref{Speed_vs_t_T=1273KtoT=973KtoT=1273K_[111]} we show the growth speed, and in Fig. \ref{Concentrations_T=1273KtoT=973KtoT=1273K_[111]}, the concentrations
profiles. First, we notice that the growth speed is fully reversible, e.g. after the temperature is quenched from 973 K to 1273 K the speed returns to its value prior 
to the cool-down. 
What is remarkable is the large factor ($\approx 40$) by which the speed increases (decreases) when the temperature is decreased (increased). 
This value can be directly compared to Fig. \ref{Speed_vs_t_and_Speed_vs_T}, which is computed at the same $P_0$ and at a \textit{constant} temperature throughout the entire growth phase. 
There, the factor by which the speed changes is 7.5 when the temperature is dropped from 1273 K to 973 K. Clearly, quenching the temperature down and then up during growth results 
in a large net increase of the growth speed (notice that the growth is completed in 1.6 s in Fig. \ref{Speed_vs_t_T=1273KtoT=973KtoT=1273K_[111]} and in 5 s 
in Fig. \ref{Speed_vs_t_and_Speed_vs_T}). Closer examination shows that this increase is attributed nearly entirely to the dimers; their concentration experiences
a way more abrupt change (compared to the concentration of the atoms) when the temperature is quenched up/down. This is shown in 
Fig. \ref{Concentrations_T=1273KtoT=973KtoT=1273K_[111]}, where the concentrations are plotted before the cool-down, after the cool-down, and after the warm-up.
Such response of the concentrations to the temperature quenches is another indicator that the dimers are primarily responsible for the experimentally observed growth kinetics.

It was determined \cite{Li,Yan,Wang} that the growth slows down with time, the more so the closer the graphene islands approach each other \cite{Wang}.
In the cited papers the Cu crystallographic surface is not identified though, it is only stated that the growth is realized on a Cu foil.
Also, since the observations are made when there is several growing islands, as is always the case, the growth slowdown may not occur were it was possible to grow a single 
island.
In our modeling, the minor decrease of the growth speed is seen for Cu[100] surface, but not for Cu[111] surface. 
However, it will be fairly straightforward to incorporate another growing island into the model, which may allow to more precisely differentiate between the growth modes 
on these Cu surfaces.  For better predictive capability it may be also necessary to include the atoms desorption term in Eq. (\ref{C_eq}) and the atoms and dimers de-attachment
rates (from the island) into Eq. (\ref{x0_eq}), along with the corresponding source terms in Eqs. (\ref{C_eq}) and (\ref{C_2_eq}). It must be noted though, that a time-resolved graphene 
growth experiments that generate a high-precision data on the growth rates, as well as the matching detailed descriptions of the plethora of the growth conditions and parameters,  
are still rare, which presents quite a challenge to further tuning the model.

\vspace{-0.5cm}
\bigskip
\noindent
\textit{Acknowledgments.}$\;$ The author acknowledges constructive discussions with V. Dobrokhotov (WKU Applied Physics Institute).

\end{document}